\documentclass[12pt]{article}
\usepackage{graphicx,amsmath,amssymb}
\usepackage{indentfirst}
\usepackage[usenames]{color}
\usepackage[colorlinks=true, urlcolor=navyblue, linkcolor=navyblue, citecolor=navyblue]{hyperref}
\usepackage{epstopdf}
\usepackage{enumerate}
\usepackage{appendix}
\usepackage{lineno}
\usepackage{setspace}

\definecolor{navyblue}{rgb}{0,0.08,0.45}
\definecolor{darkred}{rgb}{0.7,0.0,0.0}
\definecolor{darkgreen}{rgb}{0,0.6,0.2}

\newcommand{\beq}{\begin{equation}}
\newcommand{\enq}{\end{equation}}
\newcommand{\beqa}{\begin{eqnarray}}
\newcommand{\beqast}{\begin{eqnarray*}}
\newcommand{\enqa}{\end{eqnarray}}
\newcommand{\enqast}{\end{eqnarray*}}
\newcommand{\nn}{\nonumber}

\newcommand{\bec}{\begin{center}}
\newcommand{\enc}{\end{center}}
\newcommand{\beqo}{\begin{quote}}
\newcommand{\enqo}{\end{quote}}
\newcommand{\bem}{\begin{minipage}}
\newcommand{\enm}{\end{minipage}}

\newcommand{\mbf}{\mathbf}

\newcommand{\req}[1]{(\ref{#1})}

\newcommand{\half}{\textstyle \frac{1}{2}}

\newcommand{\ga}{\gamma}

\newcommand{\ep}{\epsilon}

\newcommand{\la}{\lambda}

\newcommand{\si}{\sigma}

\newcommand{\vp}{\varphi}

\newcommand{\Ga}{\Gamma}

\newcommand{\La}{\Lambda}

 \newcommand{\threehalf}{{\frac{3}{2}}}
 \newcommand{\fivehalf}{{\frac{5}{2}}}

\begin{document}

\vspace{15pt}

\begin{center}
{\huge  The Spectroscopy and Form Factors of }

\vspace{10pt}

{\huge Nucleon Resonances from Superconformal}

\vspace{10pt}

{\huge  Quantum Mechanics and Holographic QCD}

\end{center}

\vspace{15pt}

\centerline{Guy F. de T\'eramond}

\vspace{3pt}

\centerline {\it Universidad de Costa Rica, San Jos\'e, Costa Rica~\footnote{{Invited talk presented at Nucleon Resonances: From Photoproduction to High Photon Virtualities  October  12-16, 2015, Trento,  Italy\\
\href{mailto:gdt@asterix.crnet.cr}{\tt
\hspace{12pt} gdt@asterix.crnet.cr}}}}


\vspace{20pt}

\begin{abstract}

The superconformal algebraic approach to hadronic physics is used to construct a semiclassical effective theory for nucleons which incorporates essential nonperturbative dynamical features, such as the emergence of a confining scale and the Regge resonance spectrum. Relativistic bound-state equations for nucleons follow from the extension of superconformal quantum mechanics to the light front and its holographic embedding in a higher dimensional gravity theory.   Superconformal algebra has been used elsewhere to describe the connections between the light mesons and baryons, but in the present context it relates the fermion positive and negative chirality states and  uniquely determines the confinement potential of nucleons. The holographic mapping of multi-quark bound states also leads to a light-front cluster decomposition of form factors for an arbitrary number of constituents.  The remarkable analytical structure which follows incorporates the correct scaling behavior at high photon virtualities and also vector dominance at low energies.  

\end{abstract}

\newpage

\tableofcontents

\section{Introduction}

The study of the dynamics and internal structure of nucleons is an intricate problem in hadronic physics. In fact, lattice QCD calculations of the excitation spectrum of the light hadrons, and particularly nucleons,  represent a formidable task due to the enormous computational complexity beyond the leading ground state configuration~\cite{Edwards:2011jj}. On the other hand, holographic methods provide new analytical tools for the study of strongly correlated quantum systems, which are complementary to other nonperturbative approaches to strongly coupled gauge theories~\cite{Brambilla:2014jmp}. The best known example is the AdS/CFT correspondence between gravity in anti-de Sitter (AdS) five-dimensional space and  conformal field theories (CFT) in physical space-time~\cite{Maldacena:1997re}, which leads to new insights into the nonperturbative dynamics of QCD.

There is a remarkable  connection of light-front quantized theories~\cite{Dirac:1949cp, Brodsky:1997de} to gravity theory in AdS:  The Hamiltonian equations in AdS space can be precisely mapped to the relativistic semiclassical bound-state equations in the light front~\cite{deTeramond:2008ht, deTeramond:2013it}.   This connection gives an exact relation between the holographic variable $z$ of AdS space and the invariant impact light-front variable $\zeta$ in physical space-time~\cite{deTeramond:2008ht, Brodsky:2006uqa}.  This connection also implies that the light-front (LF) effective potential $U$ in the LF Hamiltonian equations, corresponds to the modification of AdS  space --described in terms of a dilaton profile $\varphi$ in the string frame.  The light-front effective potential $U$ acts on the valence state and incorporates and infinite number of multiple-component higher Fock states~\cite{Pauli:1998tf}.  To compute $U$ one must systematically express higher Fock components as functionals of the lower ones. Its actual derivation remains an unsolved problem.

Faced with the enormous complexity of nonperturbative QCD, other methods,  which encompass the essential features of the strong interaction dynamics, are needed to gain further understanding into the nature of confinement physics.  This complexity can be understood from the increase of the QCD coupling in the infrared domain, which implies that an infinite number of quark and gluons are dynamically coupled~\footnote{The QCD confinement problem is of such  complexity that it could be undecidable:  This means that it is not possible to know, starting from the fundamental degrees of freedom of the QCD Lagrangian,  whether the system is gapped or not~\cite{Cubitt:2015nat}.}. Recent progress along these lines has followed from the study of conformal quantum mechanics (QM)~\cite{deAlfaro:1976je} and its mapping to the light front,  which determines the form of the LF confinement potential $U$ Ðand thus the dilaton profile $\varphi$~\cite{Brodsky:2013ar}. This procedure allows the introduction of a scale $\sqrt \la$ in the Hamiltonian while the action remains conformal invariant~\cite{deAlfaro:1976je}.

The supersymmetric extension of conformal QM, namely superconformal quantum mechanics~\cite{Akulov:1984uh, Fubini:1984hf}, can also be mapped to the semiclassical LF effective theory~\cite{deTeramond:2014asa, Dosch:2015nwa, Dosch:2015bca}  --a one dimensional QFT, and consequently to gravity theory in AdS. This new approach to hadronic physics incorporates confinement, the appearance of a massless pion in the limit of zero-mass quarks, and the Regge excitation spectrum consistent with experimental data.  Furthermore, this framework gives remarkable connections between the light meson and nucleon spectra~\cite{Dosch:2015nwa}. It also gives predictions for the heavy-light hadron spectra, where heavy charm and bottom quark masses break the conformal invariance, but the underlying supersymmetry holds~\cite{Dosch:2015bca}~\footnote{In hadronic physics  supersymmetry is an emergent dynamical property from color $SU(3)_C$ since a diquark is in the same color representation as an antiquark, namely a $\bf \bar 3 \sim \bf 3 \times \bf 3$.}.

Following  Ref.~\cite{deTeramond:2014asa} we discuss in this article how the superconformal framework leads to relativistic bound-state equations for nucleons from the mapping to light-front physics and its embedding in a higher dimensional AdS space.  In this case, the superconformal algebra relates the nucleon positive and negative chirality states  and determines  the effective confinement potential of nucleons.  In turn, this  allows us to study the  holographic embedding of the semiclassical effective theory for nucleons. In fact, in contrast to mesons, a dilaton term in the AdS fermionic action  has no dynamical effects since it can be rotated away by a redefinition of the fermion fields~\cite{Kirsch:2006he}   and a specific Yukawa-like interaction term has to be introduced in the AdS  action to break conformal invariance~\cite{Abidin:2009hr}, but its form is left unspecified. The superconformal approach has thus the advantage that mesons and nucleons are treated on the same footing, and the confinement potential is  uniquely determined by the formalism --including additional spin-dependent  constant terms which are critical to describe the hadronic spectrum.

For a multi-quark bound state the light front invariant impact variable $\zeta$  corresponds to a system of an active quark plus an spectator cluster.  The holographic embedding is also characterized by a single variable $\zeta$ which is mapped to the AdS variable $z$~\cite{Brodsky:2006uqa}. For example, for a three quark system,  the three-body problem is reduced to an effective two-body problem where two of the constituents form a diquark cluster.  However,  in the present framework the diquark is not a tightly bound state.  It is also important to notice that the reduction to a single variable is also crucial in the superconformal formulation of hadron physics where mesons and nucleons are in the same multiplet.

The  light-front cluster decomposition of hadronic bound states is also important to solve the standing problem of  the twist assignment of the proton in holographic QCD~\cite{Brodsky:2014yha}.  Since the lowest bound-state solution to the holographic Dirac equation corresponds to twist 2, the nucleon is described by the wave function of a quark-diquark cluster. At high energies, however, all the constituents in the proton are resolved and therefore the fall-off of the form factor is governed by the number of all constituents, {\it i.e.}, it is twist 3. A related problem was found in  the study of sequential decay chains in baryons~\cite{TheCBELSA/TAPS:2015ula}, which are sensitive to the short distance behavior of the wave function. The solution to this  problem follows from the LF cluster decomposition for bound states~\cite{Brodsky:1983vf, Brodsky:1985gs, BdTDL}. It will be discussed below.

\section{Light-front holographic embedding \label{LFEMB}}

As a brief review,  we first examine the embedding  of the semiclassical light-front wave equations for mesons in AdS space. We study then the general structure of the LF equations for nucleons starting from the Dirac AdS action. The actual confinement potential is determined in the next section from the superconformal algebraic structure.

Our starting point is an effective action in AdS$_{5}$ space for the spin-$J$ tensor field {  $\Phi_{N_1  \dots N_J}$
\begin{multline}
\label{SM}
S_{\it eff} = \int d^{4} x \,dz \,\sqrt{g}  \; e^{\vp(z)} \,g^{N_1 N_1'} \cdots  g^{N_J N_J'}   \Big(  g^{M M'} D_M \Phi^*_{N_1 \dots N_J}\, D_{M'} \Phi_{N_1 ' \dots N_J'}  \\
 - \mu_{\it eff}^2(z)  \, \Phi^*_{N_1 \dots N_J} \, \Phi_{N_1 ' \dots N_J'} \Big),
 \end{multline}
with  coordinates $x^M =\left(x^\mu, z\right)$. The holographic variable is $z$ and  $x^\mu$ are Minkowski flat space-time coordinates. The metric determinant is
  $\sqrt{g} = (R/z)^{5}$ and $D_M$ is the covariant derivative which includes the affine connection ($R$ is the AdS radius). The  dilaton field $\varphi$  breaks the maximal symmetry of AdS, and the effective AdS mass $\mu_{eff}$  is determined  by the mapping to light-front physics~\cite{deTeramond:2013it}.

A hadron with momentum $P$ and physical polarization $\ep_{\nu_1 \cdots \nu_J}(P)$   is represented by 
\beq \label{scalarcov} \Phi_{\nu_1 \cdots \nu_J}(x, z) = 
e ^{ i P \cdot x} \,  \ep_{\nu_1 \cdots \nu_J}({P}) \,  \Phi_J(z) , \enq
with invariant hadron mass $P_\mu P^\mu \equiv \eta^{\mu \nu} P_\mu P_\nu = M^2$. Variation of the action \req{SM} leads to the wave equation
\beq  \label{AdSWE}
 \left[
   -  \frac{ z^{3- 2J}}{e^{\varphi(z)}}   \partial_z \left(\frac{e^{\varphi(z)}}{z^{3-2J}} \partial_z   \right)
  +  \frac{(\mu \,R )^2}{z^2}  \right]  \Phi_J = M^2 \Phi_J,
  \enq
where   $(\mu \, R)^2 = (\mu_{\it eff}(z) R)^2  - J z \, \vp'(z) + J(5 - J)$ is a constant determined by kinematical conditions in the light front~\cite{deTeramond:2013it}.

We now compare the wave equations in the dilaton-modified AdS space with LF bound-state equations in the semiclassical approximation described in~\cite{deTeramond:2008ht}. 
In the light front the hadron four-momentum  generator is $P =  (P^+, P^-, \mbf{P}_{\!\perp})$,~$P^\pm = P^0 \pm P^3$, and  the hadronic spectrum is computed from the invariant   Hamiltonian $P^2  =  P_\mu P^\mu = P^- P^+ -  \mbf{P}_\perp^2$:
\beq  \label{HLF}
P^2 \vert  \psi(P) \rangle =  M^2 \vert  \psi(P) \rangle,
\enq
where  $ {\vert \psi(P)\rangle}$ is expanded in multi-particle Fock states  $\vert n \rangle $:  ~$\vert \psi \rangle = \sum_n \psi_n \vert n \rangle$.

In the limit of zero-quark masses,  the bound-state dynamics of the constituents can be separated from the longitudinal kinematics and the orbital dependence  in the transverse LF plane leading to the wave equation~\cite{deTeramond:2008ht, deTeramond:2013it}:
\beq \label{LFWE}
\left(-\frac{d^2}{d\zeta^2}
- \frac{1 - 4L^2}{4\zeta^2} + U\left(\zeta \right) \right)
\phi(\zeta) = M^2 \phi(\zeta),
\enq
where the invariant transverse variable in impact space   $\zeta^2= x(1-x)\mbf{b}_\perp^2$ is  conjugate to a two-body LF invariant mass $M_{q \bar q}^2 = \mbf{k}_\perp^2/x(1-x)$.  The critical value of the orbital angular momentum $L = 0$ corresponds to the lowest possible stable solution~\cite{Breitenlohner:1982jf}.  Eq. \req{LFWE} is a relativistic and frame-independent LF Schr\"odinger equation: The confinement potential $U$ is instantaneous in LF time  and comprises all interactions, including those with higher Fock states.  
 Upon  the substitution  $\Phi_J(z)   \sim z^{(d-1)/2  - J} e^{-\varphi(z)/2} \, \phi_J(z)$  and ~ $z \! \to\! \zeta$   in Eq. \req{AdSWE},  we find 	Eq. \req{LFWE} with
\beq  \nn
 U(\zeta) = \half \varphi''(\zeta) +\frac{1}{4} \varphi'(\zeta)^2  + \frac{2J - 3}{2 z} \varphi'(\zeta),
\enq
and $(\mu R)^2 = - (2-J)^2 + L^2$~\cite{deTeramond:2013it, deTeramond:2010ge}. The effective LF confining potential $U(\zeta)$  thus corresponds to the IR modification of  AdS space.

Nucleons with arbitrary half-integer spin $J = T + \half$ are described in AdS by an effective action for  Rarita-Schwinger  spinors  $\Psi_{N_1 \cdots N_T}$
\begin{multline} \label{SN}
~~~  S_{\it  eff} = \half  \int  \!
d^{d} x \,dz\,  \sqrt{g} \, g^{N_1\,N_1'} \cdots g^{N_T\,N_T'}  \\
\left[ \bar  \Psi_{N_1 \cdots N_T}  \Big( i \, \Ga^A\, e^M_A\, D_M
-  \mu - \rho(z)\Big)
 \Psi_{N_1' \cdots N_T'} + h.c. \right] , ~~~~~
 \end{multline} 
where curved space indices  are $M, N = 0, \cdots 4$ and tangent indices are  $A, B = 0, \cdots, 4$, with  $e^M_A$  the inverse vielbein, $e^M_A = \left(\frac{z}{R}\right) \delta^M_A$.  The covariant derivative $D_M$  includes the affine connection and the spin connection. The tangent-space Dirac matrices obey the usual anticommutation relation $\left\{\Gamma^A, \Gamma^B\right\} = 2 \eta^{A B}$.  The effective interaction $\rho(z)$ breaks conformal symmetry and generates a baryon spectrum~\cite{Abidin:2009hr}.

A nucleon with four-momentum $P$ and chiral spinors  $u^\pm_{\nu_1 \cdots \nu_T} ({P})$ is represented by
\beq \label{Psipsi}
\Psi^\pm_{\nu_1 \cdots \nu_T}(x, z)   = e^{ i P \cdot x}  \,  u^\pm_{\nu_1 \cdots \nu_T} ({P}) \, \left(\frac{R}{z} \right)^{T-2}   \Psi^\pm_T(z).
\enq
Variation of the action \req{SN} leads to a system of linear coupled  equations
\begin{eqnarray} \label{LDE}  \nonumber
- \frac{d}{d z} \psi_-  - \frac{\nu+\half}{z}\psi_-  -  V(z) \psi_-&=& M \psi_+ , \\
 \frac{d}{d z} \psi_+ - \frac{\nu+\half}{z}\psi_+  - V(z) \psi_+ &=& M \psi_- , 
\end{eqnarray}
where  $\mu R =  \nu + \half$ ,$\psi_\pm \equiv \Psi_T^\pm$, and
\beq \label{V}
V(z) = \frac{R}{z} \, \rho(z),
\enq
a $J$-independent potential.   Thus, independently of the specific form of the potential, the value of the nucleon masses along a given Regge trajectory depends only on the LF orbital angular momentum $L$~\footnote{This result was also found in Ref. \cite{Gutsche:2011vb}.},  in agreement with the observed near-degeneracy in the baryon spectrum~\cite{Klempt:2009pi}.

Mapping to the light front, $z \to \zeta$, Eq.  \req{LDE}  is equivalent to the system of second order equations
 \beqa \label{LFWEa}
\left(-\frac{d^2}{d\zeta^2}
- \frac{1 - 4 L^2}{4\zeta^2} + U^+(\zeta) \right) \psi_+ & \! =  \! & M^2 \psi_+,  \\   \label{LFWEb}
\left(-\frac{d^2}{d\zeta^2}
- \frac{1 - 4(L + 1)^2}{4\zeta^2} + U^-(\zeta) \right) \psi_- &  \! =  \! & M^2 \psi_-,
\enqa
where
\beq \label{Upm} \nn
U^\pm(\zeta) = V^2(\zeta) \pm V'(z) + \frac{1 + 2 \nu}{\zeta} V(\zeta),
\enq
and $L = \nu = \mu R - \half$. The plus and minus component wave equations \req{LFWEa} and \req{LFWEb} correspond, respectively,  to LF orbital angular momentum $L$ and $L + 1$.

\section{Superconformal quantum mechanics and nucleon bound-state equations}

We follow Ref. \cite{deTeramond:2014asa} to construct nucleon bound-state equations  by extending  the superconformal algebraic structure of Fubini and Rabinovici~\cite{Fubini:1984hf} to the light front.   Superconformal quantum mechanics is a one-dimensional quantum field theory invariant under conformal  and supersymmetric transformations. Imposing  conformal symmetry leads to a unique choice of the superpotential  and thus to a unique confinement potential in the light front.   In addition to the Hamiltonian $H$ and the usual fermionic operators $Q$ and $Q^\dagger$ of supersymmetric quantum mechanics~\cite{Witten:1981nf}, an additional generator $S$, which is related to the generator of conformal transformations $K$, is introduced. 

We use the  representation of the operators
$$  Q   =   \chi \left(\frac{d}{d x} + \frac{f}{x} \right),   \quad
 Q^\dagger =  \chi^\dagger \left(- \frac{d}{d x} + \frac{f}{x}\right),$$
where $f$ is a dimensionless constant, and  $S = \chi \, x$,  $S^\dagger = \chi^\dagger x$. It is now simple to verify the closure of the enlarged algebraic structure. In a Pauli matrix representation:
\beqa
\half \{Q,Q^\dagger\} & \! \!= \! \!& H,    \hspace{77pt} \half \{S,S^\dagger\} = K, \nn  \\
\half \{Q ,S^\dagger\} & \! \!= \! \!& \frac{f}{2} + \frac{\si_3}{4} - i D, \hspace{20pt}  {\half}  \{Q^\dagger ,S\} = \frac{f}{2} + \frac{\si_3}{4} + i D,
\enqa
where the operators
\beqa
H & = & \frac{1}{2} \left( - \frac{d^2}{d x^2}  + \frac{f^2 -  \si_3 f}{x^2} \right), \nn \\
 D & = & \frac{i}{4} \left(\frac{d}{dx} x + x \frac{d}{d x} \right), \nn \\
  K  & = & \half x^2, 
\enqa
satisfy the conformal algebra:
\beq
[H,D]= iH, \qquad [H,K]= 2 i D, \qquad  [K,D]=-i K . 
\enq

Following Ref.~\cite{Fubini:1984hf} we define a new fermionic operator $R$, a linear combination of the generators $Q$ and $S$,
\beq
R =  \sqrt{u} \, Q + \sqrt{w} \, S,  \hspace{30pt}  R^\dagger =  \sqrt{u} \, Q^\dagger + \sqrt{w} \, S^\dagger,
\enq
which generates a new Hamiltonian $G$ 
\beq
 \half \{R, R^\dagger\}  =  G,  
\enq
where by construction
$$
\{R_\la,R_\la\} =   \{R_\la^\dagger, R_\la^\dagger\} = 0 \hspace{20pt}  {\rm and} \hspace{20pt}   [R_\la, G]  = [R_\la^\dagger, G] = 0.
$$
We find
\beq \label{Gwu}
G = u  H + w  K + \half \sqrt{u w} \,(2 f + \si_3),
\enq
which is a compact operator for   $u w > 0$.  Since the new Hamiltonian $G$ commutes with $R$ and $R^\dagger$, it  follows that  $\vert \phi \rangle$ and $R  \vert \phi \rangle$ have identical eigenvalues.

We now extend the Hamiltonian $G$  \req{Gwu} to a relativistic LF Hamiltonian by performing the substitutions
$$
x \to \zeta, \quad  f \to L + \half,  \quad \si_3 \to \ga_5, \quad 2 G \to H_{LF} \nn .
$$
We obtain:
\beqa\label{eq:LFHs}
H_{LF}  &=&  \{R, R^\dagger\}  \nn \\
& = & - \frac{d^2}{d \zeta^2} 
+ \frac{\left(L + \half\right)^2}{\zeta^2} - \frac{L +
  \half}{\zeta^2} \gamma_5 + \la^2 \zeta^2 + \la  (2 L + 1) + \la \gamma_5,
\enqa
where  $L$ is the relative LF angular momentum between the active quark and the spectator cluster and the arbitrary  coefficients $u$ and $w$ in \req{Gwu} are fixed to   $u = 1$ and $w =  \la^2$.  In a $2 \times 2$ block-matrix form the light-front Hamiltonian \req{eq:LFHs} can be expressed as
 \beq \label{HLF}
H_{LF} =  \left(  \! \begin{array}{cc}
    - \frac{d^2}{d \zeta^2}  - \frac{1 - 4 L^2}{4 \zeta^2} +  \la^2 \zeta^2 +
2 \la(L + 1) & 0\\ 
    0 & - \frac{d^2}{d \zeta^2} - \frac{1 - 4(L + 1)^2}{4 \zeta^2} + \la^2 \zeta^2 +
2 \la  L   \\  
  \end{array} \! \right).
  \enq
Since $H_{LF}$  commutes with $R$,  the eigenvalues for the chirality plus and minus eigenfunctions are identical.    Comparing \req{HLF} with \req{LFWEa} and \req{LFWEb} we obtain the effective confining potential \req{V} in AdS space: It is the linear potential $V = \la \, z$.  

The light-front eigenvalue equation $H_{LF} \vert \psi \rangle = M^2 \vert \psi \rangle$ has eigenfunctions
\begin{eqnarray} \label{psip} 
\psi_+(\zeta) &\sim& \zeta^{\frac{1}{2} + L} e^{-\la \zeta^2/2}
  L_n^L(\la \zeta^2) ,\\ \label{psim} \nn
\psi_-(\zeta) &\sim&  \zeta^{\frac{3}{2} + L} e^{-\la \zeta^2/2}
 L_n^{L+1}(\la \zeta^2), 
\end{eqnarray} 
and eigenvalues,  
\begin{equation}  \label{M2F}
M^2 = 4 \la (n + L + 1).  
\end{equation}
Both components have identical normalization~\cite{deTeramond:2014asa}:
\begin{equation} \label{eq:spnorm} \nn
\int d\zeta \,  \psi^2 _+(\zeta)   =  \int d \zeta  \, \psi^2_-(\zeta) .
\end{equation}
The nucleon spin is thus carried by the orbital angular momentum~\cite{Brodsky:2014yha}.

\begin{figure}[htdp]
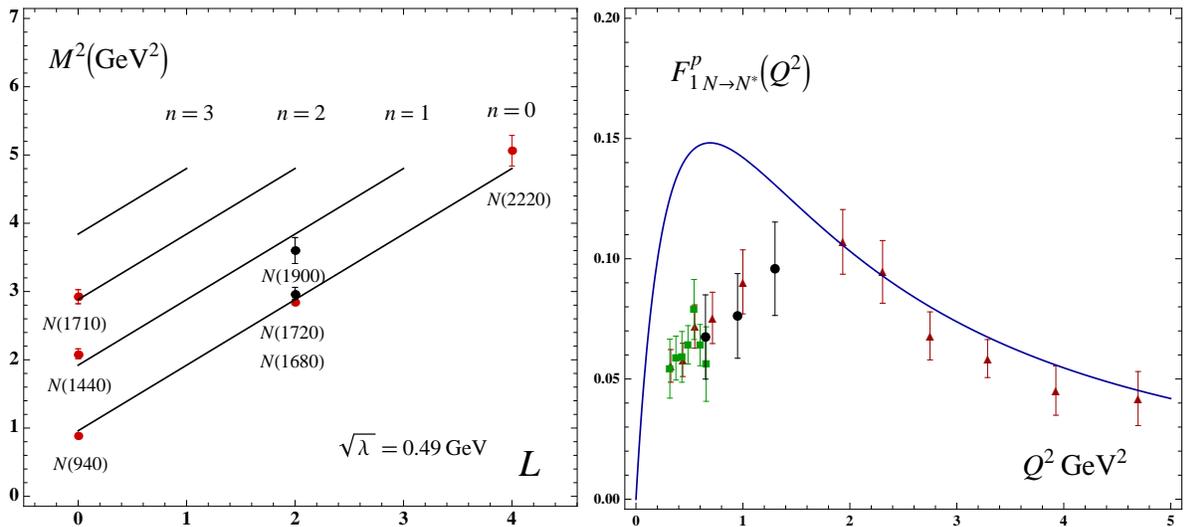

\centering
\includegraphics[width=7.6cm]{nucleon}  
\includegraphics[angle=0,width=7.9cm]{protondtff}

\caption{\small Left: Nucleon resonance spectrum of positive parity nucleons. Right: Proton transition form factor  ${F_1^p}_{N\to N^*}(Q^2)$ for the radial transition $N{\half^+}(940) \to N{\half^+}(1440)$.  Nucleon spectrum data from PDG~\cite{Agashe:2014kda}  (left)  and JLAB~\cite{Aznauryan:2009mx, Mokeev:2012vsa,  Mokeev:2015lda}  (right).}
 \label{fig}
\end{figure} 

The predictions for the resonance spectrum of the positive-parity internal spin-$\half$ light nucleons is shown in  Fig. \ref{fig} (a) for  the parent Regge trajectory,  $n =0$, and  the daughter trajectories for $n=1$, $n = 2, \cdots $. The lowest stable state, the nucleon $N{\half^+}(940)$, corresponds to $n=0$ and $L = 0$. The Roper state $N{\half^+}(1440)$ and the $N{\half^+}(1710)$ are described  as the first and second radial excited states of the nucleon. The model is also successful in explaining the $J$-degeneracy for states with the same orbital angular momentum, such as the $L = 2$ positive-parity doublet  $N{\threehalf^+}(1720) - N{\fivehalf^+}(1680)$. Only confirmed PDG~\cite{Agashe:2014kda} states are shown.

\section{Light-front cluster decomposition and form factors}

The problem of the twist assignment of the proton  in holographic QCD has been addressed in Ref. \cite{BdTDL}.  The solution to this outstanding problem is based on the LF cluster decomposition for bound states given in Refs. \cite{Brodsky:1983vf, Brodsky:1985gs},  where  the factorization properties of the deuteron were extensively analyzed. As a result of the cluster decomposition, the deuteron wave function factorizes into two distinct nucleon wave functions convoluted with a two-body factor $f_d$,  $F_d(Q^2) = f_d(Q^2) \,F_N^2( \tfrac{1}{4} Q^2)$,  where  $Q^2 f_d(Q^2) \simeq const$ at large $Q^2$.  The nucleon form factors $F_N$ are evaluated at $Q/2$, since both nucleons share the momentum transferred to the bound state by the incoming probe~\footnote{The form factors of the deuteron where studied in the framework of AdS/QCD in Ref. \cite{Gutsche:2015xva}.}.

For simplicity we examine here  spin-non-flip transition amplitudes. On the higher dimensional gravity theory it  corresponds to the  coupling of an external electromagnetic (EM) field $A^M(x,z)$  propagating in AdS with a  fermionic mode $\Psi_P(x,z)$, given by the left-hand side of the equation 
 \begin{multline} \label{FF}
 \int d^4x \, dz \,  \sqrt{g}   \,  \bar\Psi_{P'}(x,z)
 \,  e_M^A  \, \Gamma_A \, A^M(x,z) \Psi_{P}(x,z) \\  \sim 
 (2 \pi)^4 \delta^4 \left( P'  \! - P - q\right) \epsilon_\mu \bar u(P') \gamma^\mu F_1(q^2) u({P}),
 \end{multline} 
The expression on the right-hand side  represents the Dirac EM form factor in physical space-time. It is the EM  spin-conserving matrix element  of the quark current  $J^\mu = e_q \bar q \gamma^\mu q$  with local coupling to the constituents. A precise mapping of the matrix elements  can be carried out at fixed LF time, providing an exact correspondence between the holographic variable $z$ and the LF impact variable $\zeta$ in ordinary space-time~\footnote{For $n$ partons the invariant LF variable $\zeta$ is the $x$-weighted definition of the transverse impact variable of the $n-1$ spectator system~\cite{Brodsky:2006uqa}: $\zeta = \sqrt{\frac{x}{1-x}} \big\vert \sum_{j=1}^{n-1} x_j \mbf{b}_{\perp j} \big\vert$ where $x = x_n$ is the longitudinal momentum fraction of the active quark.}.

For an $N$ constituent  bound state the mapping expressed by Eq. \req{FF} leads to the analytic form~\cite{Brodsky:2014yha, Brodsky:2007hb}
\beq \label{FFN}
  F_{\tau = N}(Q^2) =  \frac{1}{{\left(1 + \frac{Q^2}{M^2_{n=0}} \right) }
 \left(1 + \frac{Q^2}{M^2_{n=1}}  \right) \cdots  \left(1 + \frac{Q^2}{M^2_{n = \tau -2}}  \right)},
 \enq 
 where the twist $\tau$ is equal to the number of constituents, {\it i. e.}, $\tau = 3$ for the proton valence state.  To compare with the data one has to shift the poles to their physical location at $-Q^2 =  4 \la(n+\half)$, with $\sqrt \la = M_\rho/2$: The predicted  bound-state poles of the $\rho$ ground state and its radial excitations~\cite{Brodsky:2014yha}.  Following this procedure we obtain the cluster form for the twist-$\tau$ elastic form factor 
\beq 
F_{\tau = N}(Q^2) =
F_{\tau = 2}(Q^2)  \,F_{\tau = 2}
(\tfrac{1}{3}Q^2)\,  \cdots\,    F_{\tau = 2}
(\tfrac{1}{2 N - 3}Q^2),
\enq 
which is the product of twist-two (pion) form factors evaluated at different scales. For the proton elastic Dirac form factor the corresponding expression is
\begin{equation}
F_1^p(Q^2)  = F_{\pi}(Q^2)\,  F_{\pi}  (\tfrac{1}{3}Q^2).
\end{equation}
It corresponds to the factorization of the proton form factor as a product of a point-like quark and composite diquark form factors with the correct scaling behavior at large $Q^2$.

A similar LF  cluster decomposition for arbitrary twist can be obtained, for example,  for  the radial transition $n = 0 \to n =1$~\cite{BdTDL}.  The Dirac transition form factor for the radial transition $N{\half^+}(940) \to N{\half^+}(1440)$ is given by  the twist-3 expression~\cite{BdTDL, deTeramond:2011qp}
\beq
F_{1\, N \to N^*}^p (Q^2)  =  \frac{ 2 \sqrt{2}}{3}\, F_{\pi \to \pi'}(Q^2)\, F_{\pi} (\tfrac{1}{5}Q^2),
\enq
which is the product of the radial transition form factor of the pion $F_{\pi \to \pi'}(Q^2)$,
\beq
F_{\pi \to \pi'}(Q^2) = \frac{1}{2}\,  \frac{Q^2/ M_\rho^2}{{\left(1 + \frac{Q^2}{M^2_\rho} \right) \left(1 + \frac{Q^2}{M^2_{\rho'}}  \right) }},
\enq
and the pion elastic form factor.  The predictions are shown in Fig. \ref{fig}~\footnote{An AdS/QCD computation including the $A^p_{1/2}$ and $S^p_{1/2}$ amplitudes is given in Ref. \cite{Gutsche:2012wb}.}. The twist-3 transition form factor behaves as  $F_{ \tau =3} \sim \frac{1}{Q^4}$ at large momentum transfer and reproduces quite well the data above $Q^2 \ge 2 ~ {\rm GeV}^2$. The low energy data, however,  is not well reproduced and possibly indicates the necessity to include higher Fock components which give important contributions at low $Q^2$, but vanish  rapidly at large $Q^2$. For example a $\tau = 5$ higher Fock component  $\vert q q q \bar q q\rangle$, in addition to the $\tau = 3$ valence   $\vert q q q \rangle$ state,  contributes to the low energy region and vanishes as $F_{ \tau =5} \sim 1/{Q^8}$ at higher energies. Indeed,  it was found for the pion form-factor that the inclusion of higher Fock components are required to have a meaningful comparison with experiment~\cite{Brodsky:2014yha}.

\section{Concluding remarks}

The connection of light-front dynamics, classical gravity  in AdS space and superconformal quantum mechanics leads to a semiclassical approximation which describes the dynamics and internal structure of nucleons. The superconformal algebraic structure  determines uniquely the effective confinement potential and connects mesons to baryons. The emerging confinement scale $\sqrt \la$  is directly related to physical observables, such as a hadron mass, and can be related to scheme dependent perturbative scales, such as the QCD renormalization scale $\La_s$~\cite{Deur:2014qfa}.  The light-front holographic mapping also leads to a cluster decomposition of hadron bound states, which is particularly useful for understanding the structure of transition amplitudes.  The analytical structure of the form factors incorporates the short distance scaling behavior dictated by the number of constituents $N$ and the transition to the nonperturbative region  determined by vector dominance.   For a hadron with angular momentum $L$, the expression \req{FFN} is still valid but the twist is $\tau = N + L$. This result could be used to test  supersymmetric QM connections at the amplitude level: It implies that the higher value of the orbital angular momentum $L$ of a partner meson, $L_M = L_B + 1$, is compensated by the additional constituent in the baryon~\cite{Dosch:2015nwa}.

Extension of the results  described here to the full light-hadron spectrum has been carried out recently by enforcing superconformal symmetry in the holographic embedding of the light-front bound-state equations, including internal spin and quark masses~\cite{BdTDL}~\footnote{A recent extension of light front-holographic QCD to describe octet and decouplet baryons is given in Ref.~\cite{Liu:2015jna}.}.	 Other relevant questions left open in Ref.~\cite{Dosch:2015nwa}, such as the nature of the spin interaction and the correct twist assignment of the baryon wave functions, have also been addressed in Ref. \cite{BdTDL}.  The new framework based on the superconformal algebraic structure and its mappings to light front holographic bound-state equations, provides a set of new analytic tools which can be particularly useful for the theoretical  interpretation of the upcoming results at the new energy scales and kinematic regions which are about to be explored with the JLab 12 GeV Upgrade Project~\cite{Aznauryan:2012ba}.

\section*{Acknowledgments}

The results presented here are based on collaborations with  Stanley J. Brodsky, Alexandre Deur, Hans Guenter Dosch and C\'edric Lorce. I
want to thank the organizers of the ECT* 2015 Workshop on Nucleon Resonances for their hospitality at Trento.

\end{document}